\documentstyle[12pt]{article}
\input psfig 
\def\be{\begin{equation}} 
\def\ee{\end{equation}}           
\def\ba{\begin{array}}  
\def\ea{\end{array}} 
\def\beqn{\begin{eqnarray}} 
\def\eeqn{\end{eqnarray}} 
\def\bt{\begin{tabular}} 
\def\et{\end{tabular}} 
\def\bc{\begin{center}} 
\def\ec{\end{center}} 
 
\def\vud{$|V_{ud}|$} 
\def\vus{$|V_{us}|$} 
\def\vcb{$|V_{cb}|$}  
 
\def\vcd{$|V_{cd}|$} 
\def\vcs{$|V_{cs}|$} 
\def\vtd{$|V_{td}|$}

\def\rub{$|\frac{V_{ub}}{V_{cb}}|$} 
\def\mu{$m_u$}

\def\sin2{sin$2\beta$} 
\def\b{$\beta$} 
\def\del{$\delta$} 
 
\begin{document} 
\title{Exploring the construction of ``Reference'' triangle 
through unitarity}
\author{Monika Randhawa and Manmohan Gupta \\ 
{\it Department of Physics,}\\ 
{\it Centre of Advanced Study in Physics,}\\ 
 {\it Panjab University, Chandigarh- 
  160 014, India.}}
  \maketitle 
\begin{abstract} 
Motivated by the possibility of the low value of \sin2~ in the  
measurements of BABAR and BELLE collaborations, we have explored 
the possibilty of construction of 
 reference unitarity triangle   
using the  unitarity of the CKM matrix, the existence of nonzero 
CP violating phase  $\delta$ and the 
experimental values of the well known CKM elements, 
without involving any  inputs from the processes which might include   
the new physics effects.  
The angles of the reference triangle are evaluated by finding 
 $\delta$ through the Jarlskog's rephasing 
invariant parameter $J$. The present data and the unitarity of the 
CKM matrix give  $\delta=50^{\rm o} \pm 20^{\rm o}$, which translates 
to $130^{\rm o} \pm 20^{\rm o}$ in the
second quadrant. The corresponding range   for \sin2~ 
 is 0.21 to 0.88. This range is 
broadly in agreement with the recently updated 
 BABAR and BELLE results. However, 
a value of \sin2$\leq$0.2, advocated by Silva and Wolfenstein 
 as a benchmark for new physics, would  
suggest a violation in the three generation unitarity and would hint 
towards the existence of a fourth generation. Further, the future  
refinements in the CKM elements will push the lower limit on \sin2~ 
still higher.
\end{abstract} 
The recent measurements of the time dependent CP asymmetry  
$a_{\psi K_S}$ in $B^o_d({\bar B}^o_d) \rightarrow \psi K_S$ 
 decay by BABAR and BELLE collaborations, for example, 
 \beqn a_{\psi K_S}& =&0.12 \pm 0.37 \pm 0.09~~~~~~~~  {\rm BABAR}~  
\cite{babar},  \label{babar} \\ 
 a_{\psi K_S}&=&0.45^{+0.43~+0.07}_{-0.44~-0.09}~~~~~~~~~~~~~~ {\rm BELLE}~  
\cite{belle}, \label{belle} \eeqn 
 look to be smaller compared to the CDF measurements \cite{cdf},  
for example, 
\be a_{\psi K_S}^{{\rm CDF}}=0.79^{+0.41}_{-0.44}~, \label{cdf} \ee 
as well as compared to the recent standard analysis of the unitarity  
triangle \cite{burasrev} with   
$|\epsilon_K|$, \rub, $\Delta m_d$ and $\Delta m_s$ 
as input, given as 
\be a_{\psi K_S}^{{\rm SM}} = 0.67 \pm 0.17. 
\label{burasrev} \ee 
 This disagreement gets reduced in the recently updated measurements 
of BABAR and BELLE, for example, 
 \beqn a_{\psi K_S}& =&0.34 \pm 0.20 \pm 0.05~~~~~~~~  {\rm BABAR}~  
\cite{babar1},  \label{babar1} \\ 
 a_{\psi K_S}&=&0.58^{+0.32~+0.09}_{-0.34~-0.10}~~~~~~~~~~~~~~ {\rm BELLE}~  
\cite{belle1}, \label{belle1} \eeqn 
however, the possibility of the $a_{\psi K_S}$ being lower than the 
predictions of Standard Model (SM) is still not ruled out.  
In the SM, $a_{\psi K_S}$  is related to the angle $\beta$ 
of the unitarity triangle as, 
\be  a_{\psi K_S} = Sin2\beta.  \label{sin2} \ee 
 
Recently,  several authors \cite{kagan} - \cite{nxb}  have 
 explored the implications of the possibility of low 
value of \sin2~ in comparison to the CDF measurements  
as well as to the global analysis of the unitarity  triangle. 
These analyses lead to the general consensus that the 
possibility of new physics could be more prominent  
in the loop dominated  processes, in particular the  
$B^o - \bar{B^o}$ mixing. Further, it is realized that the new  
physics will not affect the tree level decay processes and  
the unitarity of the three generation CKM matrix in the SM approaches 
as well as in its extensions \cite{kagan}-\cite{ut5}.  
In this connection, for better appraisal of new physics, 
it has been generally recommended  to construct a universal  
or reference unitarity triangle \cite{kagan},\cite{ut1}-\cite{ut5}, 
wherein  the inputs are free from the processes which might include   
the new physics effects, in particular the 
$B^o - \bar{B^o}$ mixing and $K^o - \bar{K^o}$ mixing parameters. 
Keeping this in mind several strategies,  
model dependent \cite{ut1,ut2} as well as model independent  
\cite{ut3,ut4,ut5}, have been formulated to construct the triangle, 
 however by and large both approaches rely on the  
rare decays. The reference triangle to be constructed is defined as, 
\be V_{ud}V_{ub}^{*} + V_{cd}V_{cb}^{*} + V_{td}V_{tb}^{*} = 0, 
\label{db} \ee 
 obtained by employing the orthogonality of the 
 first and third column of the CKM matrix (henceforth referred to as 
 triangle $db$). 
In this triangle the elements involving $t$ quark 
have not been experimentally measured as yet and hence to  
construct the   triangle, the inputs from rare decays 
involving elements $V_{td}$ and $V_{tb}$ through loops  
have to be used. 
 
In this context, it is interesting to note that 
despite several analyses of the CKM 
phenomenology in the past \cite{burasrev}, \cite{jarlskog} 
-\cite{parodi} 
 yielding valuable information,  
the implications of three generation unitarity have not been 
examined in detail  in the construction of the reference triangle. 
A reference triangle constructed purely from the considerations of 
unitarity as well as using experimentally measured CKM elements 
 will be free from the effects of new physics and hence could serve 
as a tool for deciphering deviation from the SM in measuring the 
CP asymmetries. 
 
 The purpose of the present communication is to explore
the construction of  
the  triangle $db$ using unitarity of the three generation CKM matrix 
as well as the existence of nonzero CP violating phase $\delta$, 
 by evaluating the   Jarlskog's Rephasing Invariant Parameter $J$  
and consequently the $\delta$. In particular, 
we intend to evaluate angles $\alpha$, $\beta$ and $\gamma$ 
of the triangle $db$ and study the implications of the low value of \sin2~ 
for unitarity. 
 
   To begin with, we consider the six non diagonal relations 
     implied by the unitarity of the 
CKM matrix. One of the relations corresponds to  equation 
\ref{db} and the other five are as follows, 
\beqn 
  ds~~~~~~~V_{ud}V_{us}^{*} + V_{cd}V_{cs}^{*} + V_{td}V_{ts}^{*} = 0,  
\label{ds}  \\ 
  sb~~~~~~~V_{us}V_{ub}^{*} + V_{cs}V_{cb}^{*} + V_{ts}V_{tb}^{*} = 0, 
\label{sb}  \\ 
  ut~~~~~~~V_{ud}V_{td}^{*} + V_{us}V_{ts}^{*} + V_{ub}V_{tb}^{*} = 0,   
\label{ut}  \\ 
  uc~~~~~~~V_{ud}V_{cd}^{*} + V_{us}V_{cs}^{*} + V_{ub}V_{cb}^{*} = 0,   
\label{uc}  \\ 
  ct~~~~~~~V_{cd}V_{td}^{*} + V_{cs}V_{ts}^{*} + V_{cb}V_{tb}^{*} = 0. 
\label{ct} 
\eeqn 
 The letters before the equations denote the respective  triangles. 
 
\begin{figure} 
   \centerline{\psfig{figure=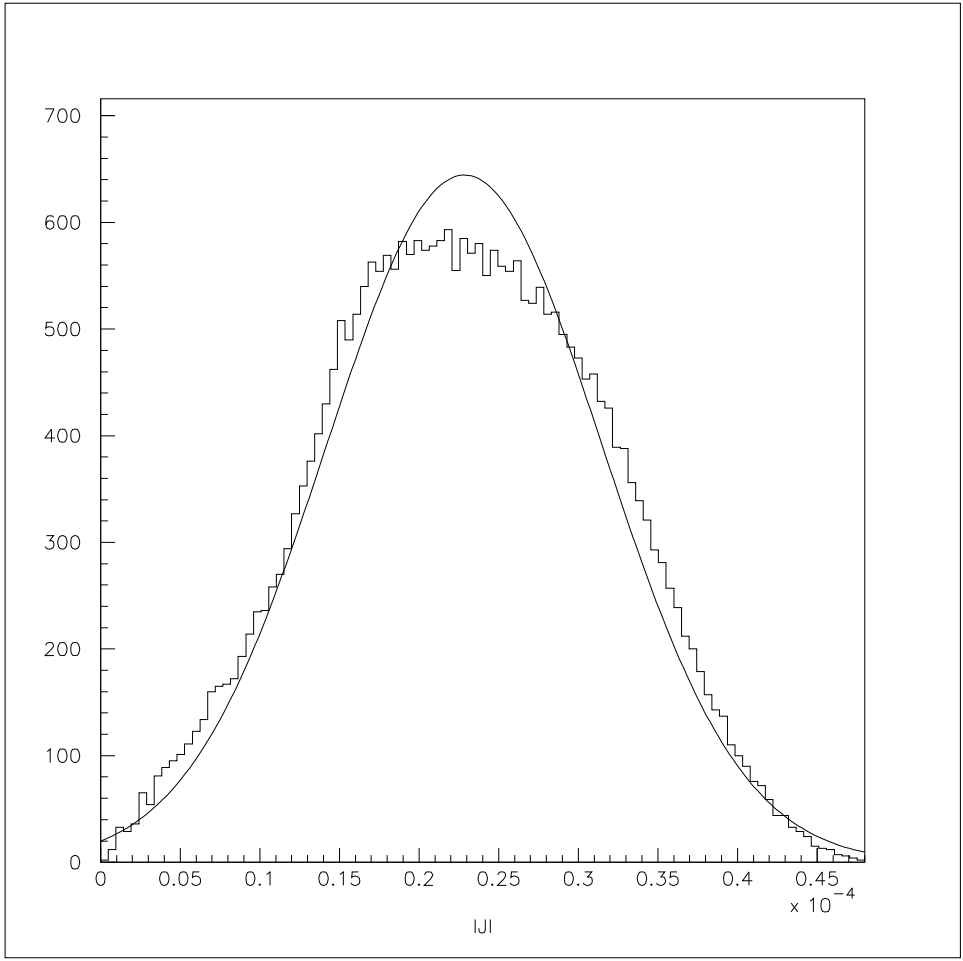,width=5in,height=5in}} 
   \caption{Gaussian fitted  to the histogram of $|J|$ plotted 
   by considering the triangle $uc$ with the input constraints 
 $|a|+|c| > |b|$ and $|b|+|c| > |a|$, where $a = |V_{ud}^*V_{cd}|$, 
  $b=|V_{us}^*V_{cs}|$ and  
  $c=|V_{ub}^*V_{cb}|$.} 
  \label{fig1} 
  \end{figure}

As mentioned above, in the triangle $db$ the elements 
 $V_{td}$ and $V_{tb}$ are not  experimentally 
measured, therefore the triangle cannot be constructed 
 without additional inputs.
 This also corresponds to our ignorance of the CP 
violating phase $\delta$ of the CKM matrix, defined as \cite{pdg}
  \be V_{CKM}= \left( \ba {lll} c_{12} c_{13} & s_{12} c_{13} & 
  s_{13}e^{-i\delta} \\ 
  -s_{12} c_{23} - c_{12} s_{23} s_{13}e^{i\delta} & 
 c_{12} c_{23} - s_{12} s_{23}s_{13}e^{i\delta} 
  & s_{23} c_{13} \\ 
  s_{12} s_{23} - c_{12} c_{23} s_{13}e^{i\delta} & 
  - c_{12} s_{23} - s_{12}c_{23} s_{13}e^{i\delta} & 
  c_{23} c_{13} \ea \right),  \label{ckm} \ee 
  with $c_{ij}=cos\theta_{ij}$ and   $s_{ij}=sin\theta_{ij}$ for  
 $i,j=1,2,3.$ 
In the above representation, the mixing angles $s_{12}$, $s_{23}$ 
and $s_{13}$  can be obtained   from the experimentally well known elements 
\vus, \vcb~  and 
 \rub~  given in Table 
 \ref{tabinput}, hence the CP violating phase 
 $\delta$ remains the only unknown parameter in determining the triangle 
 $db$. The phase $\delta$, however, is related to the  
Jarlskog's rephasing invariant  parameter $J$ as
      \be J = s_{12}s_{23}s_{13}c_{12} 
        c_{23}c^2_{13}sin \delta. \label{j} \ee 
Therefore, an evaluation of $J$ would allow us to find $\delta$ and  
consequently the angles  $\alpha$, \b~ and $\gamma$ of the  
 triangle $db$. To evaluate $J$, we make use of the fact  
that the areas of all the six triangles (equations \ref{db}-\ref{ct}) 
are  equal and that the area of any of the unitarity triangle 
 is related to Jarlskog's Rephasing Invariant Parameter $J$ as 
\be J = 2 \times {\rm Area~ of~any~ of~ the~ Unitarity~ Triangle.} 
 \label{area} \ee 
This, therefore affords an opportunity to evaluate $J$ 
through one of the unitarity triangle whose sides are  
experimentally well known, for example, triangle $uc$.   
  The triangle $uc$ though is quite well known, but is highly  
squashed, therefore one needs to be careful while 
evaluating $J$ through this triangle. 
The sides of the triangle  represented by $|V_{ud}^*V_{cd}|~(=a)$ and 
  $|V_{us}^*V_{cs}| ~(=b)$ are of comparable lengths while the third side 
  $|V_{ub}^*V_{cb}| ~(=c)$  is several orders of 
    magnitude smaller compared to $a$ and $b$. 
  This creates complications for evaluating the area of the triangle 
  without violating the existence of CP violation. 
  To avoid these complications, without violating the 
unitarity, we have  incorporated the constraints 
   $|a|+|c| > |b|$ and $|b|+|c| > |a|$ \cite{branco}, 
this ensures that the triangle exists and has nonzero area. 
    Using these  constraints and the experimental data given in the 
     table \ref{tabinput}, a histogram can be plotted,  
shown in figure \ref{fig1}, to which a gaussian is  
fitted yielding the result, 
     \be J = (2.59 \pm 0.79) \times 10^{-5} 
          \label{jpdg1s}.  \ee 
 Calculating $s_{12},~ s_{23}$ and $s_{13}$ from 
    the experimental values of  \vus, \rub,~ 
    and  \vcb~ given in table \ref{tabinput}, one can plot a
 distribution for $\delta$ as well using equation \ref{j} and
\ref{jpdg1s} and incorporating the above mentioned constraints.
Again the distribution for \del~ is gaussian and yields,
\be \delta = 50^{\rm o} \pm 20^{\rm o}, \label{del1} \ee
which in the second quadrant translates to,
\be \delta = 130^{\rm o} \pm 20^{\rm o}. \label{del2} \ee
  
       This value of $\delta$ apparently looks to be the consequence 
  only of the unitarity relationship given by equation \ref{uc}. 
    However on further investigations, as shown by Branco and Lavoura 
  \cite{branco}, one finds that this $\delta$ range is consequence 
   of all the non trivial unitarity constraints. In this sense the above 
  range could be attributed to  unitarity of the 
 CKM matrix. It needs to be noted that with the above range of $\delta$ 
 and  the  experimental  values of \vus, \vcb~ and   
\rub~ given in Table \ref{tabinput}, the CKM matrix thus evaluated 
 is in excellent agreement with  PDG  CKM matrix \cite{pdg}. In
 particular  \vtd, the element with the most sensitive dependence on
\del~ comes out to lie in the range,
       \be V_{td}=0.0046 ~{\rm to}~ 0.0134, \label{vtd} \ee
in comparison with the PDG range, $V_{td}=0.004 ~{\rm to}~ 0.014.$

 In order to reinforce our conclusion regarding $J$, we
have also evaluated $J$ through other unitarity triangles given in 
equations \ref{db} to \ref{ct}.
Without getting into the details, 
which will be published elsewhere, we
have considered the evaluation of $J$ through the two unsquashed triangles 
$db$ and $ut$. Since these triangles involve elements which are 
 not experimentally known, for them we have used their PDG values
based on the unitarity considerations. Following the procedure used
for evaluating $J$ through triangle $uc$, we get the 
 following values for $J$ evaluated through 
$db$ and $ut$ respectively,
\beqn J &=& (2.51 \pm 0.87) \times 10^{-5},   \label{jdb} \\
 J &=& (2.45 \pm 0.91) \times 10^{-5}.    \label{jut}  \eeqn
 Quite interestingly, we find that
$J$ values evaluated through these triangles are very much in agreement
with the $J$ evaluated through the 
triangle $uc$. We, therefore, would like to emphasize that our 
evaluation of $J$ and $\delta$ from the squashed triangle $uc$ is
very much  consistent with the unitarity based evaluation of 
CKM matrix by PDG.
 
 After having obtained $\delta$, the triangle $db$ 
 can be constructed, however  without involving inputs from 
 the phenomena which may have influence from the new physics as 
 well as without the inputs from the rare decays. 
The angles $\alpha$, \b~ and $\gamma$ of the triangle can be 
expressed  in terms of the CKM elements as,   
\be \alpha = arg\left(\frac{-V_{td}V_{tb}^*}{V_{ud}V_{ub}^*} \right),  
\label{alpha} \ee 
\be \beta = arg\left(\frac{-V_{cd}V_{cb}^*}{V_{td}V_{tb}^*}\right), 
 \label{beta} \ee 
\be \gamma = arg\left(\frac{-V_{ud}V_{ub}^*}{V_{cd}V_{cb}^*}\right), 
 \label{gamma} \ee 
where CKM elements are as given by the PDG representation in the 
equation \ref{ckm}. In Table \ref{tabinput} we  
have listed the experimental values of the 
CKM elements as given by PDG \cite{pdg} as well as their future values. 
 Making use of the PDG representation of CKM matrix given in 
equation \ref{ckm}, experimental values of  \vus,~ \vcb~ 
and \rub~ from table \ref{tabinput} and  
\del ~ given by equations \ref{del1} and \ref{del2}, one can easily find out  
the corresponding ranges for the three angles. 
In the Table \ref{tab1}, we have listed the corresponding results for  
$J$, $\delta$,  $\alpha$, \b~ and $\gamma$. 
The  ranges for $\alpha$, \b~ and $\gamma$ are as follows, 
\beqn \alpha& \simeq &19^{\rm o}~ {\rm to}~ 139^{\rm o} 
 \label{alphauni}, \\ 
 \beta& \simeq &6^{\rm o} ~{\rm to} ~ 31^{\rm o} \label{betauni}, \\ 
 \gamma &\simeq &30^{\rm o} ~{\rm to} ~ 70^{\rm o}~~~~~
{\rm in~ I~ quadrant, ~~~ and} \\
 & &   110^{\rm o} ~{\rm to} ~ 150^{\rm o}~~~~~
{\rm in ~II~ quadrant}. \label{gammauni} \eeqn 
While evaluating the three angles, we have taken care that the triangle 
is closed. 
The range of  \sin2~ corresponding to equation \ref{betauni} is given as, 
 \be sin2\beta =  0.21~ {\rm to} ~0.88 \label{sin2uni}. \ee 
It needs to be emphasized that this range for \sin2~ is obtained by  
making use of unitarity and the well known CKM elements listed 
 in Table \ref{tabinput}. The above range has considerable overlap 
with the BABAR and BELLE results,  
however if \sin2~ is found to be $\leq$0.2, a benchmark for new physics 
as advocated by Silva and Wolfenstein \cite{silva}, then one may  
conclude that even the three generation unitarity may not be valid and 
one may have to go to four generations to explain the low values of \sin2. 
In such a scenario, the widely advocated assumption \cite{kagan} 
-\cite{ut5} that the non SM physics 
resides in loop dominated processes only may not be valid. 
  
 A few comments are in order. 
It needs to be pointed out that while evaluating the area of the 
 unitarity triangle $uc$, we have assumed that CP violating phase 
 $\delta$ is nonzero and that the triangle $uc$ exists and hence 
 has nonzero area. In order to incorporate this in plotting the  
histogram we have not considered the entries corresponding to area 
 of the triangle being zero and the case corresponding to one of the 
sides being larger than the sum of the other two sides. Both of these 
possibilities are not unambiguously ruled out by the present data because of  
the fact that the uncertainties in the two larger sides are greater 
than the third shorter side. 
 However, the important point to be noted is the fact that our evaluation
of $J$ and $\delta$ obtained by incorporating the above mentioned
constraints does reproduce the PDG CKM elements based on the
experimental input and unitarity. 

It is interesting to examine the consequences of the future refinements  
in  the CKM elements.  
While listing the future values of the elements we have considered only 
 those elements where the present error is more than 15$\%$, 
for example \rub~ and \vcs. The future values of these 
elements are listed in column III of Table  
\ref{tabinput}. 
One finds from the Table \ref{tab1} that the refinements in  \rub~  
and \vcs~   would improve the lower bound on \sin2~ from 0.21 to 0.31. 
This would give a clear signal for physics beyond the SM  
in case \sin2~ is measured to be $\leq$ 0.2. To emphasize this conclusion, 
we have also considered all the future inputs at their 90$\%$ CL  
and this gives the lower limit of \sin2=0.18. 
 
It may be of interest to mention that  recent investigations 
involving texture 4 zeros quark mass matrices  
and unitarity \cite{massmat},  
yield the following range for \sin2,  
\be Sin2\beta = 0.27~ {\rm to}~ 0.60, \label{massmat} \ee 
which looks to be compatible with the present unitarity based  
calculations. A value of \sin2 $\leq$ 0.2 therefore, will have far 
 reaching consequences for unitarity as well as for texture 
 specific mass matrices \cite{massmat,massmat1,ito}. 
 
It is interesting to compare our results (equation \ref{sin2uni}) with 
those of Buras  (equation \ref{burasrev}), obtained  
from the measurements of $|\epsilon_K|$, \rub, $\Delta m_d$  
and $\Delta m_s$, which look to be much 
narrower compared to ours. This is easy to understand when one considers  
the definition of \b~ given in equation \ref{beta}, wherein the magnitude  
and phase of $V_{td}$ play an important role. For example, the  
range of $\delta$ given by equation \ref{del1} and 
\ref{del2} yields the 
$V_{td}$ range as 0.0046 to 0.0134, whereas the  
range corresponding to Buras's analysis is 0.0067 to 0.0093, 
which is narrower 
primarily due to restrictions imposed by $|\epsilon_K|$, $\Delta m_d$  
and $\Delta m_s$. Our preliminary investigations wherein we have incorporated 
the constraints due to  $|\epsilon_K|$, $\Delta m_d$  
and $\Delta m_s$ along with the unitarity lead to results 
which are in agreements with those of Buras.
 
To conclude, we have explored the possibility of construction 
of a reference unitarity triangle by making  
use of the three generation unitarity of the CKM matrix, 
the existence of nonzero 
CP violating phase  $\delta$ and the 
experimental values of the well known CKM elements, 
without involving any  inputs from the processes which might include   
the new physics effects, in particular the 
$B^o - \bar{B^o}$ mixing and $K^o - \bar{K^o}$ mixing parameters 
as well as the rare decays. 
The angles of the triangle have been 
 evaluated by finding the CP violating phase $\delta$ through the  
Jarlskog's rephasing invariant parameter $J$. Also, the $J$ and 
$\delta$ evaluated through the triangle $uc$ lead to the magnitudes
of CKM matrix elements which are in full agreement with the
 PDG CKM matrix.
The present data and the unitarity of the 
CKM matrix give  $\delta=50^{\rm o} \pm 20^{\rm o}$, which translates 
to $130^{\rm o} \pm 20^{\rm o}$ in the
second quadrant. The corresponding range   for \sin2~ 
 is 0.21 to 0.88. This range is
broadly in agreement with the recently updated 
 BABAR and BELLE results and 
also has considerable overlap with the range found from the 
texture 4 zeros quark mass matrices and the unitarity of the  
CKM matrix. However, 
a value of \sin2$\leq$0.2 advocated by Silva and Wolfenstein 
 as a benchmark for new physics would  
imply a violation in the three generation unitarity and would hint 
towards the existence of a fourth generation. Further, the future  
refinements in the CKM elements will push the lower limit on \sin2~ 
still higher, for example from 0.21 to 0.31, thus 
 giving a clear signal for physics beyond the SM  
in case \sin2~ is measured to be $\leq$ 0.2. This remains valid 
even when the future values are considered at their 90$\%$ CL. 
 \vskip 1cm 
  {\bf ACKNOWLEDGMENTS}\\ 
M.G. would like to thank S.D. Sharma for useful discussions. 
M.R. would like to thank CSIR, Govt. of India, for 
 financial support and also the Chairman, Department of Physics, 
for providing facilities to work in the department. 
                     
\newpage 
 
\begin{table}  
\bc \begin{tabular}{|l|l|l|} \hline 
Parameter & PDG values \cite{pdg} & Future values  \\ \hline 
 \vud & 0.9735 $\pm$ 0.0008 & 0.9735 $\pm$ 0.0008 \\  
\vus &  0.2196 $\pm$ 0.0023 & 0.2196 $\pm$ 0.0023 \\ 
\vcd & 0.224 $\pm$ 0.016 &  0.224 $\pm$ 0.016 \\ 
 \vcs & 1.04 $\pm$ 0.16 & 1.04 $\pm$ 0.08 \\ 
 \vcb & 0.0402 $\pm$0.0019 &  0.0402 $\pm$0.0019\\ 
 \rub &  0.090 $\pm$ 0.025 & 0.090 $\pm$ 0.010 \\  
 & & \\ \hline 
\end{tabular} 
\caption{Values of the CKM parameters used throughout the paper.} 
\label{tabinput} 
\ec \end{table} 
 
\begin{table} 
\bc 
\begin{tabular}{|c|c|c|c|} \hline 
&  With PDG values  &  
 With future values &  \bt{c} With future values \\ at their 
 90$\%$ CL \\ \et \\ \hline 
& & & \\ 
 $J$ & $ (2.59 \pm 0.79) \times 10^{-5} $ & 
$(2.79 \pm 0.49) \times 10^{-5} $ & $ (2.61 \pm 0.78) 
 \times 10^{-5}$ \\ & & & \\ 
 $\delta$ & \bt{c} 50$^{\rm o} \pm 20^{\rm o}$, \\
130$^{\rm o} \pm 20^{\rm o}$ \\ \et &  
\bt{c} 60$^{\rm o} \pm 18^{\rm o}$, \\
120$^{\rm o} \pm 18^{\rm o}$ \\ \et   &
\bt{c} 55$^{\rm o} \pm 20^{\rm o}$, \\
125$^{\rm o} \pm 20^{\rm o}$ \\ \et \\
 &  & & \\  
$\alpha$ & $19^{\rm o}$ to $141^{\rm o}$ & 
$28^{\rm o}$ to $124^{\rm o}$ & $19^{\rm o}$ to $143^{\rm o}$ \\ & & & \\ 
 $\beta$ & $ 6^{\rm o}$ to $31^{\rm o}$  &  
 $9^{\rm o}$ to $31^{\rm o}$ & $5^{\rm o}$ to $36^{\rm o}$ \\ & & & \\
$\gamma$ & \bt{c} 50$^{\rm o} \pm 20^{\rm o}$, \\
130$^{\rm o} \pm 20^{\rm o}$ \\ \et &  
\bt{c} 60$^{\rm o} \pm 18^{\rm o}$, \\
120$^{\rm o} \pm 18^{\rm o}$ \\ \et   &
\bt{c} 55$^{\rm o} \pm 20^{\rm o}$, \\
125$^{\rm o} \pm 20^{\rm o}$ \\ \et \\ \hline 
\end{tabular}  
\caption{$J$, $\delta$ and corresponding $\alpha$, $\beta$ 
and $\gamma$ with PDG and the future values of  
input parameters listed in Table 1} 
\label{tab1} 
\ec \end{table} 

\newpage  
 
\end{document}